\documentclass{article}

\usepackage{microtype}
\usepackage{graphicx}
\graphicspath{{images/}}
\usepackage{booktabs}

\usepackage[preprint]{icml2026}

\usepackage{mathtools}
\usepackage{amssymb}

\usepackage{hyperref}
\usepackage[capitalize,noabbrev]{cleveref}

\icmltitlerunning{Probabilistic Modeling of Venture Capital Portfolio Outliers}

\begin{document}

\twocolumn[
  \icmltitle{Probabilistic Modeling of Venture Capital Portfolio Outliers}

  \begin{icmlauthorlist}
    \icmlauthor{Kensei Sakamoto}{yyy}
    \icmlauthor{Hasan Ugur Koyluoglu}{bbb}
    \icmlauthor{Fuat Alican}{zzz}
    \icmlauthor{Yigit Ihlamur}{zzz}
  \end{icmlauthorlist}

  \icmlaffiliation{yyy}{University of Oxford}
  \icmlaffiliation{zzz}{Vela Research}
  \icmlaffiliation{bbb}{Oliver Wyman}

  \icmlcorrespondingauthor{Kensei Sakamoto}{kensei.sakamoto@trinity.ox.ac.uk}
   \icmlcorrespondingauthor{Yigit Ihlamur}{yigit@vela.partners}

  \vskip 0.3in
  \noindent\textbf{Abstract.}
  In this paper, we define probabilistic measures for venture portfolio performance based on individual outlier probability for each investment and the dependence across investments. This work is inspired by loan portfolio modeling against default risk used in banking. In mathematical terms, we calculate the probability distribution of the sum of N non-homogeneous Boolean outcomes (investments becoming outliers) that are correlated through common factors such as overall market conditions and sector effects. Specifically, we implemented a latent-factor model in which each investment’s success is the exceedance of a Gaussian latent variable composed of idiosyncratic returns and returns from interpretable shared factors (stock markets, industry sector indices, geography and founder type). The formulation follows a simulation approach to preserve heterogeneous deal-level success probabilities and uses empirically estimated correlation matrices.  When applied to synthetic portfolios, our model reveals that expected outlier counts alone are insufficient statistics for evaluating venture portfolios. Portfolios with identical expected outcomes can exhibit drastically different levels of reliability and risk (such as having no outlier in N investments) when various levels and forms of correlation are embedded. Ceteris paribus, diversification improves the probability of achieving minimum number of outliers by reducing exposure to common shocks such as the same industry. However, this improvement comes at the cost of lower upside, underscoring a fundamental tradeoff between reliability and magnitude of clustered outlier successes. By illustrating how portfolio composition, size, and factor exposures interact with correlation to shape the probabilities of zero, one, and multiple outliers, the framework provides a practical, testable bridge between deal-level outlier probability assessment and objective-aware portfolio construction. In light of these results and modeling assumptions, limitations, data-proxy choices, and extensions (including time-varying correlations and heavier-tailed latent factors) are discussed. 
  \vskip 0.3in
]

\printAffiliationsAndNotice{}

\vspace{1em}

\section{Introduction}

It is well established that venture capital returns are dominated by a small number of extreme outcomes \citep{hallwoodward2010}. These exceptionally successful deals, often referred to as \emph{unicorns}, account for a disproportionate share of a fund's total performance, while the majority of investments fail to produce sufficient returns. As a result, venture portfolio performance depends not on mean returns, but rather on the realization of rare outlier successes. This feature makes venture investing fundamentally different from traditional asset classes and renders expectation values insufficient for assessing portfolio quality.

A key challenge in constructing venture portfolios is that investment outcomes are rarely independent. Companies that share sectoral focus, geographic exposure, or founder characteristics are exposed to common economic and technological shocks. These shared exposures induce correlation across outcomes, increasing the likelihood of joint successes and failures. Despite this, common heuristic portfolio construction rules such as diversification or concentration are often justified without a formal treatment of the underlying dependence between deals.

In this paper, we propose a flexible and interpretable framework for modeling correlated venture outcomes. We adopt a latent-variable approach inspired by credit risk models \citep{vasicek1987, Koyluoglu1998}, in which each investment's outcome is modeled as the exceedance of a Gaussian latent variable. The variable is composed of both idiosyncratic factors and shared latent shocks associated with sector, geography, and founder characteristics. This formulation allows us to model correlation between investments while maintaining heterogeneous success probabilities at the deal level.

The remainder of the paper is organized as follows. Section 2 introduces the model framework. Section 3 describes the estimation of the correlation structure using real-world data. Section 4 examines the effects of correlation. Section 5 compares various heuristic portfolio selection choices under correlation. Finally, Section 6 concludes and provides a discussion of limitations and future work.

\section{Model Framework}
Here, we describe the general framework for modeling correlated venture outcomes. The objective is to preserve deal-level success probabilities while introducing interpretable dependence that arises from shared sectoral, geographic, and founder-type exposures.

\subsection{Latent Variable Representation}
We model each company, $i$, using the latent variable $A_i$, which represents the company's latent return index. Its exceedance past a certain threshold thus corresponds to the company becoming a unicorn. Specifically, we define
\begin{equation}
    A_i=\textbf{w}_i^\top \textbf{Z}+\sqrt{1-\textbf{w}_i^\top\boldsymbol{\Sigma}\textbf{w}_i}\epsilon_i
\end{equation}
where $\textbf{Z}\sim\mathcal{N}(0,\boldsymbol{\Sigma})$ is a vector of factor returns, $\epsilon_i\sim\mathcal{N}(0,1)$ is an idiosyncratic shock independent across companies, and $\textbf{w}_i$ is a vector of factor loadings. By construction, $A_i\sim\mathcal{N}(0,1)$.

Company $i$ is classified as a unicorn if
\begin{equation}
    A_i>\Phi^{-1}(1-p_i)
\end{equation}
where $p_i$ denotes the standalone probability of success and $\Phi$ is the standard normal cumulative distribution function. This formulation preserves individual success probabilities while introducing correlation through shared latent factors.

\subsection{Factor Loadings}

In this model, we assume that each company belongs to a set of categorical groups, including a sector, a geographic region, and a founder type. We encode these affiliations using a vector, $\textbf{r}_i$, with nonzero components indicating which groups company $i$ belongs to. 

To control the relative contribution of each group type, we assign weights $S$, $G$, and $F$ to sector, geography, and founder type, respectively, with $S+G+F=1$. In this paper, we set $S=0.6$, $G=0.3$, and $F=0.1$, reflecting the view that sectoral exposure is the dominant source of shared risk, followed by geography and founder type. These values should be interpreted as modeling priors rather than calibrated parameters.

For a company affiliated with sector $s$, geography $g$, and founder type $f$, the corresponding components of $\textbf{r}_i$ are set to $\sqrt{S}$, $\sqrt{G}$, and $\sqrt{F}$. We then define
\begin{equation}
    \textbf{b}_i=\frac{\textbf{r}_i}{\sqrt{\textbf{r}_i^\top\boldsymbol{\Sigma} \textbf{r}_i}}
\end{equation}
which ensures $\textbf{b}_i^\top \boldsymbol{\Sigma} \textbf{b}_i=1$ so that every company has unit variance prior to scaling. The final loading vector is then given by
\begin{equation}
    \textbf{w}_i=w_0\textbf{b}_i
\end{equation}
where $w_0$ controls the overall strength of dependence.

In this model, we choose $w_0$ such that the average pairwise correlation across a representative universe of companies matches a target value of 0.12. This is inspired by analogous correlation magnitudes used in credit risk modeling \citep{gordy2003}, though its applicability in the context of VC should be viewed as a modeling prior. 

To ensure this, consider the correlation between two companies, $i$ and $j$
\begin{equation}
    \rho_{ij}=\textbf{w}_i^\top\boldsymbol{\Sigma} \textbf{w}_j=w_0^2\textbf{b}_i^\top\boldsymbol{\Sigma} \textbf{b}_j=w_0^2\rho'_{ij}
\end{equation}
where we have defined $\rho'_{ij}=\textbf{b}_i^\top\boldsymbol{\Sigma} \textbf{b}_j$. Averaging over all company pairs, we get $\bar{\rho}_{ij}=w_0^2\bar{\rho}'_{ij}$ and thus we choose
\begin{equation}
    w_0=\sqrt{\frac{0.12}{\bar{\rho}'_{ij}}}
\end{equation}
Note that to ensure model validity, we require $w_0^2<1$ or the construction of the latent variable, $A_i$, breaks down.

\subsection{Simulation Procedure}
For a given portfolio with specified success probabilities and group affiliations, portfolio outcomes can be generated using Monte Carlo simulations. For each iteration, values for $\textbf{Z}$ and $\epsilon_i$ are sampled to compute the latent variables $A_i$, from which unicorn outcomes are determined via threshold exceedance. Repeating this procedure for a sufficient number of iterations yields a distribution of unicorn outcomes.

To efficiently sample correlated factor shocks according to $\textbf{Z}\sim\mathcal{N}(0,\boldsymbol{\Sigma})$, we perform a Cholesky decomposition on the correlation matrix $\boldsymbol{\Sigma}=\textbf{L}\textbf{L}^\top$ and construct 
\begin{equation}
    \textbf{Z}=\textbf{L}\textbf{X}
\end{equation}

where $\textbf{X}\sim\mathcal{N}(0,\textbf{I})$. This ensures that
\begin{equation}
    \mathrm{Var}(\textbf{Z})=\textbf{L}\mathrm{Var}(\textbf{X})\textbf{L}^\top=\textbf{L}\textbf{L}^\top=\boldsymbol{\Sigma}
\end{equation}

\section{Estimating the Correlation Matrix}
To model the dependence between latent factor groups, we require an estimate of the correlation matrix, $\boldsymbol{\Sigma}$. Since comprehensive private-market data on venture outcomes are not publicly available, for the purposes of this paper we construct an empirical correlation matrix using public equity returns. 

Specifically, we compute an empirical correlation matrix based on monthly log returns from January 2020 to December 2025. Sector factors are proxied using sector-specific exchange-traded funds, while geographic and founder-type factors are proxied using baskets of publicly traded firms selected to broadly reflect the corresponding groups.

\begin{figure}[ht]
  \vskip 0.2in
  \begin{center}
    \centerline{\includegraphics[width=\columnwidth]{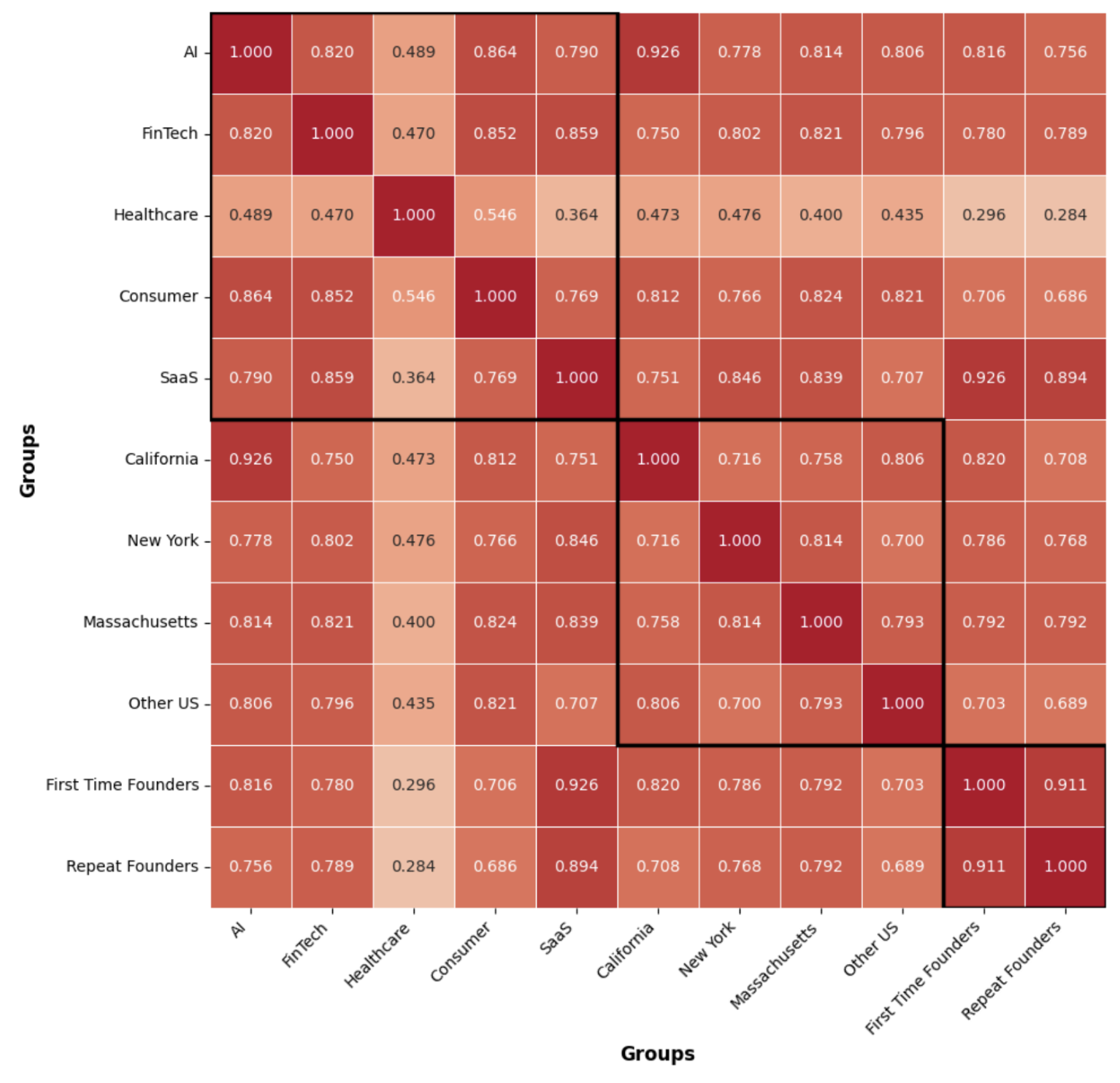}}
    \caption{
      Correlation matrix for AI, FinTech, healthcare, consumer, SaaS, California, New York, Massachusetts, other US, first-time founders, and repeat founders
    }
    \label{correlation-matrix}
  \end{center}
\end{figure}

The returns of each proxy are aggregated into a single data matrix, from which we compute the correlation matrix $\boldsymbol{\Sigma}$. In our implementation, the matrix includes five sectors, four geographic regions, and two founder types, yielding a total of 11 groups. The resulting correlation matrix is visualized in \cref{correlation-matrix}.

Several observations can be made from the estimated matrix. In particular, the healthcare sector exhibits a relatively weak correlation with the other groups, suggesting exposure to a distinct set of economic and technological factors. Given the sampling period chosen, it is also likely that the COVID-19 pandemic has heightened this effect. In contrast, the correlation between first-time and repeat founders is relatively high. This can be explained by the fact that both founder types are exposed to similar underlying market dynamics despite differences in their baseline success probabilities, leading to substantial co-movement in outcomes. Furthermore, we also observe a strong correlation between the AI sector and California, consistent with the geographic concentration of AI-focused firms within that state.

Before proceeding, it is important to address a few limitations of this approach. The main concern is whether public equity returns are really an appropriate data source for a model that is concerned with private venture outcomes. While public equity returns are an imperfect proxy for private venture outcomes, they provide an informative prior for shared economic exposures that plausibly affect both markets \citep{sorensenjagannathan2015}. In practical applications, a more robust implementation would ideally rely on historical private market valuation data if this can be obtained. Nevertheless, the constructed matrix provides a reasonable and transparent basis for exploring how correlation affects venture portfolio outcomes.

\section{Impact of Correlation}
In this section, we examine how correlation affects venture portfolio outcomes. We begin by comparing correlated and uncorrelated models under homogeneous probability assumptions, and then perform sensitivity analysis across a range of success probabilities. More general probability assumptions are explored in Section 5.

\subsection{Correlation versus Independence}

We first compare the outcomes produced from the latent-factor model with those obtained under the assumption that all outcomes are independent. For simplicity, we initially consider a single-factor model in which only sectors are correlated with each other. Furthermore, we assume that all deals have a homogeneous standalone probability of success of 4\%.

\begin{figure}[ht]
  \vskip 0.2in
  \begin{center}
    \centerline{\includegraphics[width=\columnwidth]{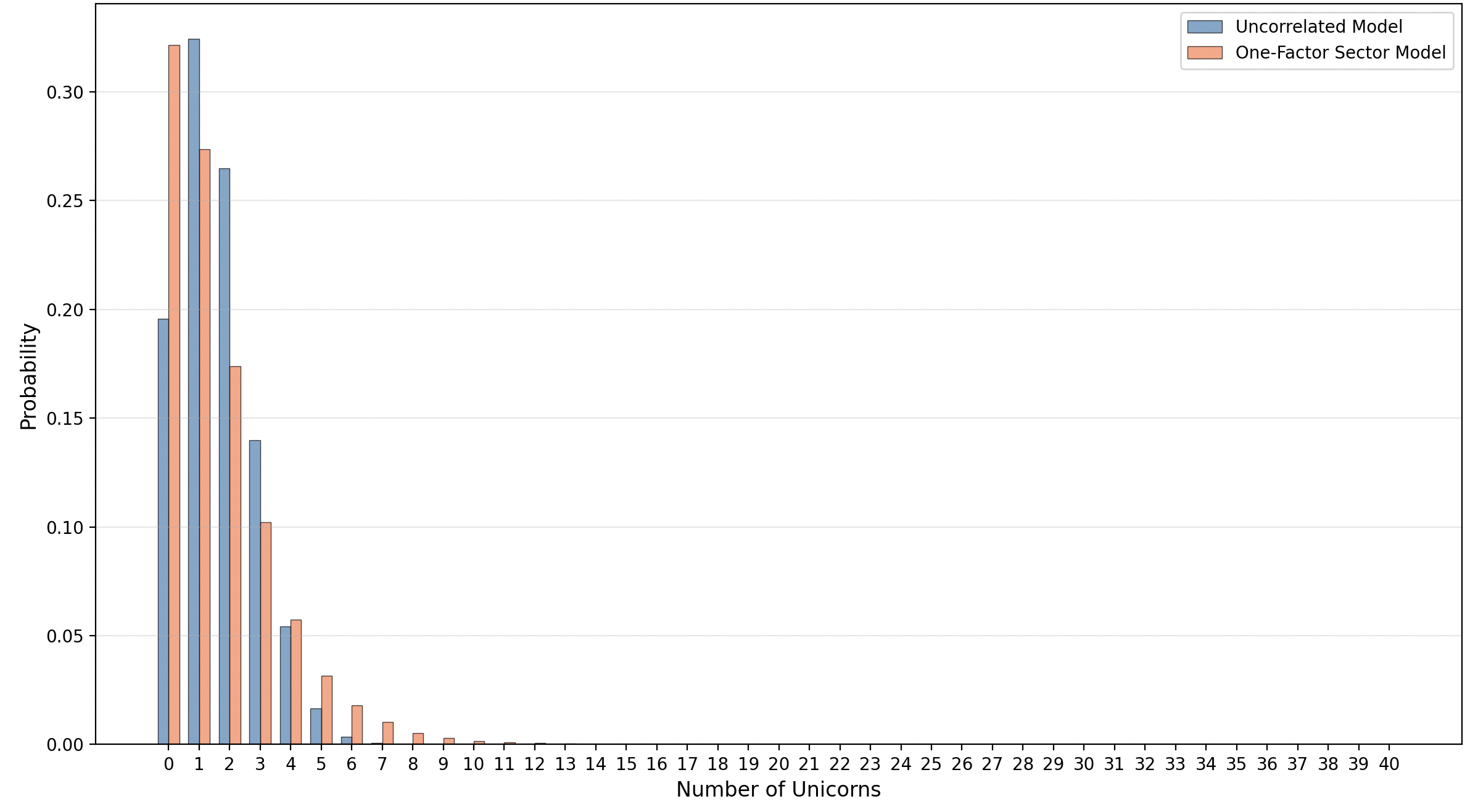}}
    \caption{
      Unicorn distribution of 40 deal portfolio with homogeneous 4\% probability of success for the uncorrelated and single-factor model
    }
    \label{uncorrelated-single-factor-distribution}
  \end{center}
\end{figure}

\cref{uncorrelated-single-factor-distribution} shows the resulting unicorn count distributions for a portfolio of 40 identical deals under the uncorrelated and single-factor models, with summary statistics reported in \cref{uncorrelated-single-factor-table-1}.

\begin{table}[t]
  \caption{Statistics comparing the uncorrelated and single-factor models for portfolio with identical deals}
  \label{uncorrelated-single-factor-table-1}
  \begin{center}
    \begin{small}
      \begin{sc}
        \begin{tabular}{lcc} 
          \toprule
           & Uncorrelated & Single-Factor \\ 
          \midrule
          $\mathbb{E}[U]$ & 1.6 & 1.6 \\ 
          $\mathbb{P}(U=0)$ & 19.6\% & 32.4\% \\
          $\mathbb{P}(U\leq1)$ & 52.0\% & 59.5\% \\
          $\mathbb{P}(U\leq2)$ & 78.5\% & 76.9\% \\
          \bottomrule
        \end{tabular}
      \end{sc}
    \end{small}
  \end{center}
  \vskip -0.1in
\end{table}

We first observe that, by construction, the expected number of unicorns is identical across both models, reflecting the fact that individual deal-level success probabilities are preserved. Despite this, the shape of the distributions slightly differs, with the introduction of correlation increasing the likelihood of joint success and failure. In particular, the probability of producing no unicorns increases from 19.6\% under independence to 32.4\% under the single-factor model.

On the other hand, since the uncorrelated distribution peaks around one unicorn, the probability of producing at most two unicorns is instead higher under independence than under correlation. This highlights that the assumption of independence can both severely underestimate and overestimate risk in venture outcomes depending on the performance threshold of interest.

Next, we consider the effect of equally diversifying the portfolio across all five sectors. The resulting statistics are shown in \cref{uncorrelated-single-factor-table-2}.

\begin{table}[t]
  \caption{Statistics comparing the uncorrelated and single-factor models for sector diversified portfolio}
  \label{uncorrelated-single-factor-table-2}
  \begin{center}
    \begin{small}
      \begin{sc}
        \begin{tabular}{lcc} 
          \toprule
           & Uncorrelated & Single-Factor \\ 
          \midrule
          $\mathbb{E}[U]$ & 1.6 & 1.6 \\ 
          $\mathbb{P}(U=0)$ & 19.6\% & 29.2\% \\
          $\mathbb{P}(U\leq1)$ & 52.0\% & 57.8\% \\
          $\mathbb{P}(U\leq2)$ & 78.5\% & 76.8\% \\
          \bottomrule
        \end{tabular}
      \end{sc}
    \end{small}
  \end{center}
  \vskip -0.1in
\end{table}

We see that while the results for the uncorrelated model do not change, the left-tail risk is reduced for the single-factor model, decreasing the probability of the portfolio failing. This illustrates the motivation for diversification in portfolio construction, as the goal in venture investing is often the realization of a minimum number of unicorns.

Finally, we extend the analysis to the full multi-factor model, incorporating sectoral, geographic, and founder-type correlations. As an example, consider the same fully sector diversified portfolio discussed previously. A comparison of the single-factor model and the multi-factor model is shown in \cref{single-factor-multi-factor}.

\begin{table}[t]
  \caption{Statistics comparing the single-factor and multi-factor models for sector diversified portfolio}
  \label{single-factor-multi-factor}
  \begin{center}
    \begin{small}
      \begin{sc}
        \begin{tabular}{lcc} 
          \toprule
           & Single-Factor & Multi-factor \\ 
          \midrule
          $\mathbb{E}[U]$ & 1.6 & 1.6 \\ 
          $\mathbb{P}(U=0)$ & 29.2\% & 32.5\% \\
          $\mathbb{P}(U\leq1)$ & 57.8\% & 59.8\% \\
          $\mathbb{P}(U\leq2)$ & 76.8\% & 76.9\% \\
          \bottomrule
        \end{tabular}
      \end{sc}
    \end{small}
  \end{center}
  \vskip -0.1in
\end{table}

Since the full model now introduces correlation between geography and founder types, there is further downside risk associated with the portfolio even if the portfolio appears to be diversified across sectors. This underscores the importance of modeling multiple channels of dependence, as diversification along one axis may be insufficient when other shared exposures remain.

\subsection{Probability Sensitivity Analysis}
To assess whether the effects documented above depend on the choice of success probabilities, we perform a sensitivity analysis in which portfolio composition is held fixed while standalone probabilities are varied. Specifically, we consider a portfolio of 40 identical deals with homogeneous success probabilities of 2\%, 4\%, 8\%, and 16\%. The corresponding results for the uncorrelated model and the full model are shown in \cref{uncorrelated-sensitivity} and \cref{full-model-sensitivity}, respectively.

\begin{table}[t]
  \caption{Probability sensitivity analysis of uncorrelated model}
  \label{uncorrelated-sensitivity}
  \begin{center}
    \begin{small}
      \begin{sc}
        \begin{tabular}{lcccc} 
          \toprule
           & 2\% & 4\% & 8\% & 16\% \\ 
          \midrule
          $\mathbb{E}[U]$ & 0.8 & 1.6 & 3.2 & 6.4 \\ 
          $\mathbb{P}(U=0)$ & 44.5\% &19.6\% & 3.7\% & 0.1\% \\
          $\mathbb{P}(U\leq1)$ & 81.0\% & 52.0\% & 16.1\% & 0.9\% \\
          $\mathbb{P}(U\leq2)$ & 95.4\% & 78.5\% & 36.8\% & 3.6\% \\
          \bottomrule
        \end{tabular}
      \end{sc}
    \end{small}
  \end{center}
  \vskip -0.1in
\end{table}

\begin{table}[t]
  \caption{Probability sensitivity analysis of full model}
  \label{full-model-sensitivity}
  \begin{center}
    \begin{small}
      \begin{sc}
        \begin{tabular}{lcccc} 
          \toprule
           & 2\% & 4\% & 8\% & 16\% \\ 
          \midrule
          $\mathbb{E}[U]$ & 0.8 & 1.6 & 3.2 & 6.4 \\ 
          $\mathbb{P}(U=0)$ & 54.6\% & 33.4\% & 14.0\% & 3.1\% \\
          $\mathbb{P}(U\leq1)$ & 80.7\% & 60.1\% & 32.2\% & 9.3\% \\
          $\mathbb{P}(U\leq2)$ & 91.7\% & 77.1\% & 49.6\% & 18.1\% \\
          \bottomrule
        \end{tabular}
      \end{sc}
    \end{small}
  \end{center}
  \vskip -0.1in
\end{table}

As expected, increasing the standalone probability of success leads to a proportional increase in the expected number of unicorns in both models. However, the evolution of the associated risk differs significantly between the two models.

\begin{figure}[ht]
  \vskip 0.2in
  \begin{center}
    \centerline{\includegraphics[width=\columnwidth]{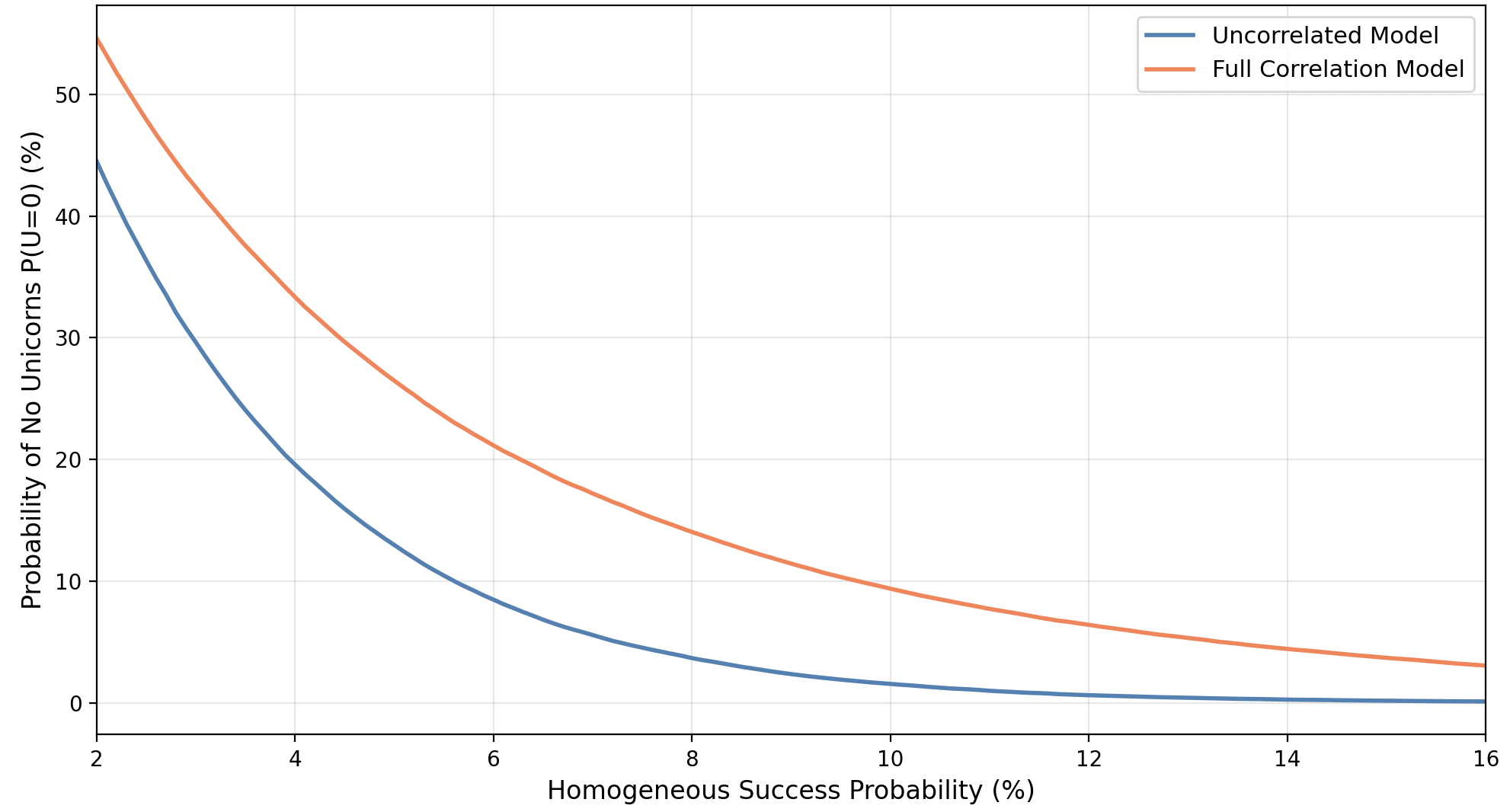}}
    \caption{
      Probability of producing no unicorns with respect to homogeneous success probability for uncorrelated and full model
    }
    \label{probability-sensitivity-graph}
  \end{center}
\end{figure}

Under independence, increases in success probability rapidly reduce the downside risk. For example, increasing the success probability from 2\% to 4\% reduces the probability of not producing unicorns from 44.5\% to 19.6\%. In contrast, under the correlated model, the risk decreases substantially more gradually, with the probability of not producing unicorns only decreasing from 54.6\% to 33.4\% for the same increase in success probability. This is visualized for the $\mathbb{P}(U=0)$ case in \cref{probability-sensitivity-graph}.

It can be observed that the uncorrelated model approaches zero more rapidly compared to the full model as the homogeneous success probability increases. These results demonstrate that correlation fundamentally limits the extent to which improvements in deal quality can reduce downside risk. This implies that increasing standalone success probabilities is not as beneficial for guaranteeing reliable portfolio performance and that managing the dependence between investments is equally critical.

\section{Portfolio Comparisons}
Having examined the effects of correlation under simplified probability assumptions, we now apply the full model to a set of stylized portfolio compositions.

\subsection{Assigning Synthetic Probabilities}
In practice, deals in a venture portfolio exhibit heterogeneous success probabilities that are typically estimated through detailed analysis. For the purposes of this paper, we assign synthetic success probabilities to each deal based on rules that broadly reflect empirical patterns observed in real venture outcomes.

Specifically, companies founded by first-time founders are assigned success probabilities drawn from a scaled Beta distribution on $[0.1\%, 12\%]$ with a target mean of 1.8\%, while those founded by a repeat founder are on $[1\%,20\%]$ with a target mean of 2.6\%. We use Beta distributions with parameters $\alpha=1$, $\beta=6$ for first-time founders and $\alpha=1$, $\beta=11$ for repeat founders, producing strongly right-skewed distributions consistent with the heavily tailed nature of venture outcomes.

To incorporate geographical and sectoral effects, we apply a +1\% nudge to companies headquartered in California or New York, or belonging to sectors commonly regarded as high-growth, such as AI, FinTech, and SaaS. Furthermore, all probabilities are capped at their respective upper bounds.

\subsection{Baseline Portfolio Comparison}
To summarize portfolio outcomes, we report statistics that capture downside risk and conditional upside. In particular, we consider the following quantities:

\begin{itemize}
  \item $\mathbb{P}(U=0)$, $\mathbb{P}(U\leq 1)$, and $\mathbb{P}(U\leq 2)$, the probabilities that a portfolio has at most 0, 1, and 2 unicorns.
  \item $\mathbb{E}[U|U\geq 1]$, $\mathbb{E}[U|U\geq 2]$, and $\mathbb{E}[U|U\geq 3]$, the expected number of unicorns conditional on having at least 1, 2, and 3 unicorns.
\end{itemize}

We begin by considering the following portfolios with fixed size of 40 deals:

\begin{itemize}
    \item Portfolio A: An industry average portfolio.
    \item Portfolio B: A fully concentrated portfolio on AI, California, and repeat founders.
    \item Portfolio C: A fully diverse portfolio across all groups.
    \item Portfolio D: A selective but diversified portfolio.
\end{itemize}

The compositions of these portfolios are summarized in \cref{baseline-composition} and their resulting outcomes are reported in \cref{baseline-comparison}.

\begin{table}[t]
  \caption{Composition of baseline portfolios}
  \label{baseline-composition}
  \begin{center}
    \begin{small}
      \begin{sc}
        \begin{tabular}{lcccc} 
          \toprule
           & A & B & C & D \\ 
          \midrule
          Repeat founders & 30\% & 100\% & 50\% & 100\%\\ 
          First-time founders & 70\% & 0\% & 50\% & 0\% \\
          AI & 30\% & 100\% & 20\% & 35\%\\
          FinTech & 15\% & 0\% & 20\% & 32.5\%\\
          Healthcare & 15\% & 0\% & 20\% & 0\%\\
          Consumer & 15\% & 0\% & 20\% & 0\%\\
          SaaS & 25\% & 0\% & 20\% & 32.5\%\\
          CA & 40\% & 100\% & 25\% & 50\%\\
          NY & 20\% & 0\% & 25\% & 50\%\\
          MA & 10\% & 0\% & 25\% & 0\%\\
          Other US & 30\% & 0\% & 25\% & 0\%\\
          \bottomrule
        \end{tabular}
      \end{sc}
    \end{small}
  \end{center}
  \vskip -0.1in
\end{table}

\begin{table}[t]
  \caption{Performance comparison of baseline portfolios}
  \label{baseline-comparison}
  \begin{center}
    \begin{small}
      \begin{sc}
        \begin{tabular}{lcccc} 
          \toprule
           & A & B & C & D \\ 
          \midrule
          $\mathbb{P}(U=0)$ & 37.0\% & 28.6\% & 37.5\% & 28.0\%\\ 
          $\mathbb{P}(U\leq1)$ & 65.5\% & 54.6\% & 65.8\% & 54.1\% \\
          $\mathbb{P}(U\leq2)$ & 81.8\% & 72.4\% & 82.2\% & 72.3\%\\
          $\mathbb{E}[U|U\geq1]$ & 2.14 & 2.58 & 2.13 & 2.57\\
          $\mathbb{E}[U|U\geq2]$ & 3.09 & 3.49 & 3.06 & 3.46\\
          $\mathbb{E}[U|U\geq3]$ & 4.06 & 4.46 & 4.05 & 4.42\\
          \bottomrule
        \end{tabular}
      \end{sc}
    \end{small}
  \end{center}
  \vskip -0.1in
\end{table}

There are a few insights that emerge from these results.

We first observe that portfolios A and C exhibit higher risk and lower conditional upside than the other portfolios, reflecting their lower exposure to repeat founders, which reduced the overall probability of attaining unicorns. Given the low weighting assigned to the founder factor and the relatively high correlation between the two founder types, the reduction in correlation achieved by diversifying founder types is dominated by the loss in average success probability.

If we now compare portfolios B and D, we find that the probability of having at most 0, 1, and 2 unicorns is marginally lower in portfolio D. While the differences are small in magnitude, they are consistent with the intuition that reduced correlation through diversification decreases the likelihood of joint failure. On the other hand, the conditional expectation values exhibit slightly lower values for the more diversified portfolio despite $\mathbb{E}[U]$ being the same. This pattern is consistent with the idea that more concentrated portfolios tend to realize larger clusters of success conditional on avoiding joint failure. Taken together, these results illustrate a fundamental tension between the two desirable quantities. Lower correlation improves the reliability of clearing a minimum success threshold, while higher correlation increases the magnitude of success once that threshold is cleared.

\subsection{Limits of Complete Diversification}

A natural question is whether complete diversification across groups with similar success probabilities necessarily minimizes left-tail risk. To investigate this nuance, we temporarily classify healthcare as a high-growth sector, assigning it the same +1\% nudge as AI, FinTech, and SaaS. This choice is motivated by the weak correlation that healthcare has with other sectors.

Now consider portfolios E, F, and G, with identical founder and geographic compositions as portfolio D but with different sector allocations shown in \cref{EFG-composition}. The corresponding results are reported in \cref{EFG-comparison}.

\begin{table}[t]
  \caption{Sector allocations for portfolios E, F, and G}
  \label{EFG-composition}
  \begin{center}
    \begin{small}
      \begin{sc}
        \begin{tabular}{lcccr} 
          \toprule
           & E & F & G \\ 
          \midrule
          AI & 25\% & 40\% & 20\%\\
          FinTech & 25\% & 20\% & 20\%\\
          Healthcare & 25\% & 20\% & 40\%\\
          SaaS & 25\% & 20\% & 20\%\\
          \bottomrule
        \end{tabular}
      \end{sc}
    \end{small}
  \end{center}
  \vskip -0.1in
\end{table}

\begin{table}[t]
  \caption{Performance comparison of portfolios E, F, and G}
  \label{EFG-comparison}
  \begin{center}
    \begin{small}
      \begin{sc}
        \begin{tabular}{lccc} 
          \toprule
           & E & F & G \\ 
          \midrule
          $\mathbb{P}(U=0)$ & 27.45\% & 27.53\% & 27.38\%\\
          $\mathbb{P}(U\leq1)$ & 53.69\% & 53.80\% & 53.64\%\\
          $\mathbb{P}(U\leq2)$ & 72.12\% & 72.20\% & 72.09\%\\
          \bottomrule
        \end{tabular}
      \end{sc}
    \end{small}
  \end{center}
  \vskip -0.1in
\end{table}

We first observe that portfolios E and F exhibit nearly identical downside risk, with portfolio F showing a marginally higher probability of poor outcomes, consistent with its increased concentration in the AI sector. In contrast, portfolio G exhibits slightly lower downside risk than portfolio E, despite a comparable reduction in sectoral diversity, reflecting the weak correlation of healthcare with other sectors.

This result highlights a critical point: even among groups with similar success probabilities, the portfolio that minimizes correlation is not necessarily the most diversified one. Effective diversification thus requires awareness of the underlying correlation structure.

\subsection{Portfolio Size Sensitivity Analysis}

\begin{figure}[ht]
  \vskip 0.2in
  \begin{center}
    \centerline{\includegraphics[width=\columnwidth]{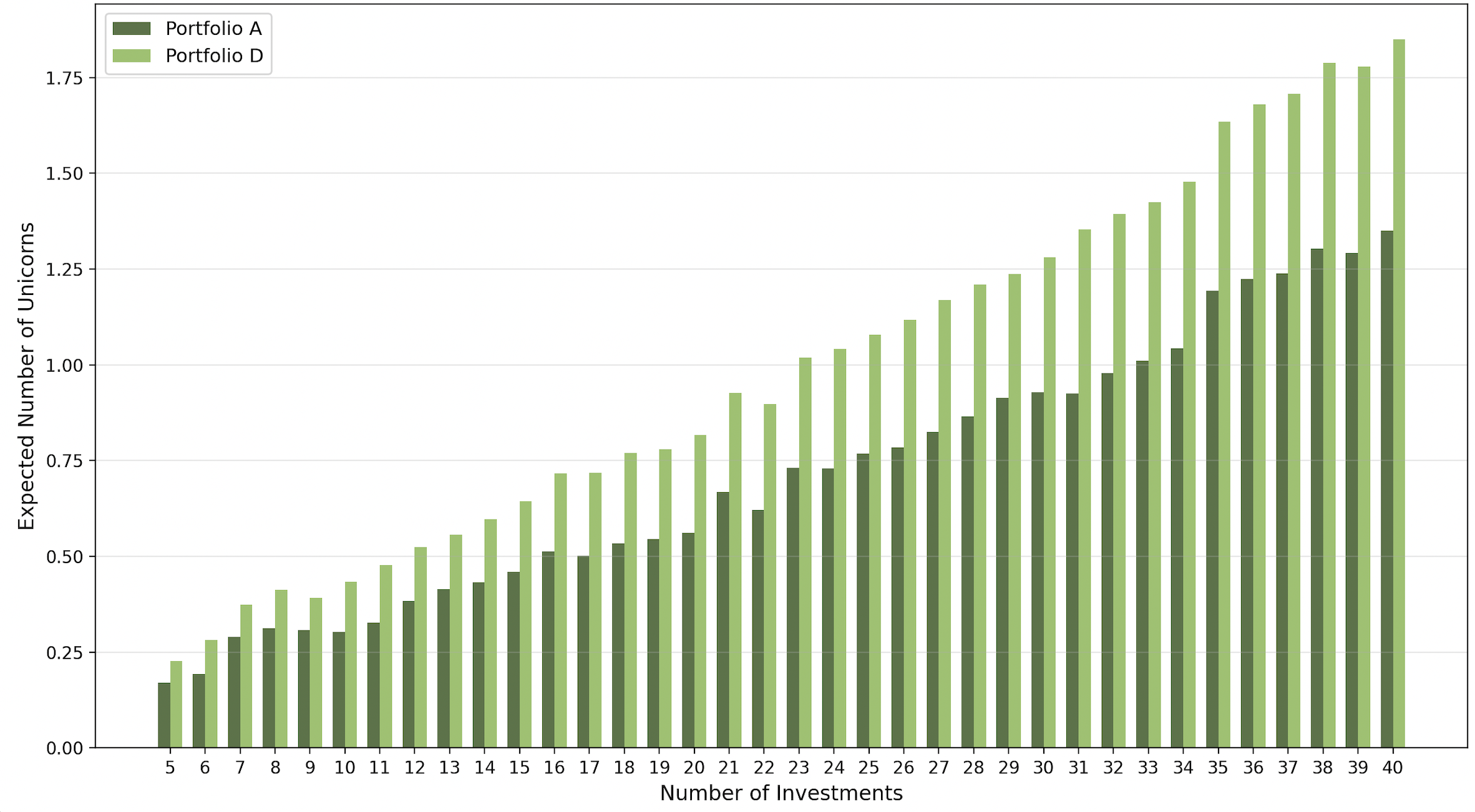}}
    \caption{
      Expected number of unicorns by number of investments for portfolios A and D
    }
    \label{expected-unicorns-sensitivity}
  \end{center}
\end{figure}

\begin{figure}[ht]
  \vskip 0.2in
  \begin{center}
    \centerline{\includegraphics[width=\columnwidth]{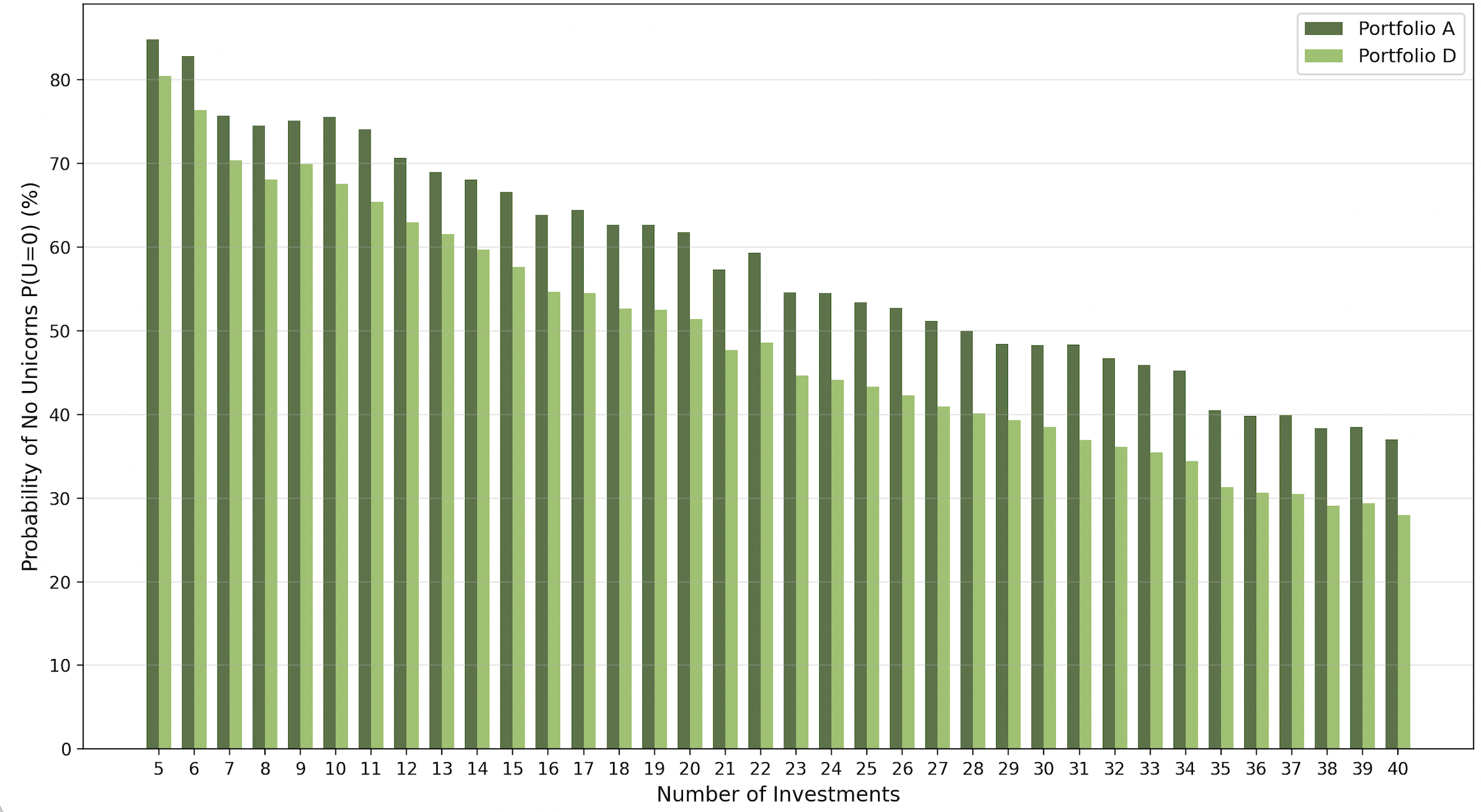}}
    \caption{
      Probability of no unicorns by number of investments for portfolios A and D
    }
    \label{probability-no-unicorns-sensitivity}
  \end{center}
\end{figure}

Next, we examine how portfolio size affects the corresponding outcomes. Holding the portfolio composition fixed, we vary the number of investments from 5 to 40 and compare the industry average portfolio (Portfolio A) to the selective but diversified portfolio (Portfolio D). For each size, we report both the expected number of unicorns and the probability of returning no unicorns. The results are shown in \cref{expected-unicorns-sensitivity} and \cref{probability-no-unicorns-sensitivity}.

As expected, \cref{expected-unicorns-sensitivity} shows that the expected number of unicorns increases approximately linearly with respect to the portfolio size in both cases. Moreover, the selective portfolio consistently exhibits higher expected number of unicorns across all sizes due to its higher average standalone success probabilities. More importantly, the gap in expected unicorns between the two portfolios widens as portfolio size increases, reflecting the cumulative effect of higher-quality deal selection.

In contrast, \cref{probability-no-unicorns-sensitivity} reveals a slightly different behavior. First, since the probability of producing no unicorns must converge to zero as portfolio size grows, reductions in risk display diminishing returns as opposed to linear improvement. Furthermore, while the more selective portfolio consistently exhibits a lower probability of no unicorns than the industry average portfolio, the magnitude of this advantage increases only gradually with respect to portfolio size. This contrasts sharply with the behavior observed with the expected unicorn count and suggests that increasing portfolio size is more effective in amplifying expected outcomes than in mitigating risk.

\section{Conclusion}
In this paper, we introduced a latent-factor framework for modeling correlated venture capital outcomes and compared various heuristic portfolio selection methods. By representing each investment’s success as the exceedance of a latent Gaussian variable with shared sectoral, geographic, and founder-type exposures, the model captures realistic dependence structures while preserving heterogeneous deal-level success probabilities.

\subsection{Key Takeaways}

Our analysis highlights several key insights. 

First, expected unicorn counts alone are insufficient statistics for evaluating venture portfolios. Portfolios with identical expected outcomes can exhibit drastically different levels of reliability and risk in the presence of correlation.

Second, diversification can improve the probability of achieving minimum success thresholds by reducing exposure to common shocks. However, this improvement comes at the cost of lower conditional upside once those thresholds are met, underscoring a fundamental tradeoff between reliability and magnitude of clustered success. Importantly, portfolios that minimize correlation are not necessarily the most diversified in a simplistic sense, and allocations toward weakly correlated groups can be beneficial even when those groups have lower standalone success probabilities. 

Third, while increasing the number of investments scales expected outcomes, it offers diminishing returns in reducing left-tail risk under correlation even for selective and diversified portfolios. As a result, even relatively large portfolios remain vulnerable to common economic shocks that can lead to joint failure of deals.

Fourth, correlation fundamentally limits the extent to which improvements in deal-level success probabilities can reduce downside risk. Even substantial increases in standalone success probabilities lead to only gradual reductions in left-tail risk when outcomes are correlated, implying that deal quality improvements alone are insufficient to guarantee reliable portfolio performance.

Finally, optimal venture portfolio construction depends critically on the investor’s objective. Portfolios that minimize the probability of failing to reach a minimum success threshold are not the same as those that maximize conditional upside once success occurs, highlighting the need for objective-aware portfolio design under dependence.

More broadly, the proposed framework provides a tractable way to stress-test venture portfolios under realistic dependence structures, offering a bridge between heuristic portfolio construction and formal risk modeling. The framework establishes a flexible foundation that can be applied for determining optimal portfolio structures in future research.

\subsection{Limitations and Future Work}
One key limitation in our analysis is the assumption that each company belongs to only one sector, which is often not the case. Extending the model to allow for multi-sector affiliations would provide a more realistic representation of venture portfolios, but will substantially increase model complexity and likely reduce interpretability.

For a similar reason, our analysis largely abstracts from the interaction between sector, geography, and founder type at the individual deal level. While we find that increasing exposure to repeat founders is generally beneficial under the assumed correlation structure and probability assignment, this conclusion may not hold if repeat founder deals are concentrated in specific sectors or geographies. Accounting for such interactions would be important for more rigorous analysis, but lies beyond the scope of this paper.

Furthermore, while we only consider sector, geography, and founder type as sources of shared risk, venture outcomes could be influenced by other latent factors as well which could be explored in future work.

Turning to the model itself, a primary concern, as mentioned earlier, is the use of public equity proxies to estimate the correlation between the various groups. While public markets may reflect shared underlying economic factors, they are likely to be an imperfect substitute for private market dynamics. Future work could incorporate private market valuation data or funding round outcomes to obtain a more reliable estimate for dependence across venture investments.

Another limitation is the assumption that the correlation between the various groups is static over time. In reality, venture markets are cyclical, with periods of increased systemic risk followed by more idiosyncratic periods. In a more robust model, we would want to enable the correlation matrix to evolve over time, allowing the model to encapsulate different periods in a venture cycle.

Finally, following common practice in latent-factor credit risk models, the framework assumes a Gaussian specification for the latent factors. However, given the heavy-tailed nature of venture outcomes, this assumption may underestimate the likelihood of joint extreme clustering. Extending the framework to incorporate heavier-tailed latent factors, such as Student-t distributions, would allow the model to better capture extreme co-movement in venture outcomes.

\bibliographystyle{apalike}
\bibliography{references}

\end{document}